\documentclass[a4paper,11pt]{article}
\usepackage{a4wide}
\usepackage{amsthm}
\newtheorem{theorem}{Theorem}
\newtheorem{lemma}{Lemma}
\usepackage[utf8]{inputenc}
\usepackage[T1]{fontenc}

\usepackage{amsmath}
\usepackage{graphicx}
\usepackage{tikz}
\usetikzlibrary{arrows}
\usetikzlibrary{decorations.pathmorphing}
\pgfmathsetmacro{\constantx}{1.2}
\pgfmathsetmacro{\constanty}{1.2}
\pgfmathsetmacro{\nodescale}{0.60}
\pgfmathsetmacro{\letterfigsc}{0.75}
\pgfmathsetmacro{\letterscale}{1}

\DeclareMathOperator{\level}{level}

\title{Subsequence Automata with Default Transitions}

\author{Philip Bille \and Inge Li G\o rtz \and Frederik Rye Skjoldjensen}
\date{\small Technical University of Denmark\\ 
  \footnotesize\texttt{\{phbi,inge,fskj\}@dtu.dk}}

\begin{document}
\maketitle
\begin{abstract}
Let $S$ be a string of length $n$ with characters from an alphabet of size $\sigma$. The  \emph{subsequence automaton} of $S$ (often called the \emph{directed acyclic subsequence graph}) is the minimal deterministic finite automaton accepting all subsequences of $S$. A straightforward construction shows that the size (number of states and transitions) of the subsequence automaton is $O(n\sigma)$  and that this bound is asymptotically optimal. 

In this paper, we consider subsequence automata with \emph{default transitions}, that is, special transitions to be taken only if none of the regular transitions match the current character, and which do not consume the current character. We show that with default transitions, much smaller subsequence automata are possible, and provide a full trade-off between the size of the automaton and the \emph{delay}, i.e., the maximum number of consecutive default transitions followed before consuming a character. 

Specifically, given any integer parameter $k$, $1 < k \leq \sigma$, we present a subsequence automaton with default transitions of size $O(nk\log_{k}\sigma)$ and delay $O(\log_k \sigma)$. Hence, with $k = 2$ we obtain an automaton of size $O(n \log \sigma)$ and delay $O(\log \sigma)$. On the other extreme, with $k = \sigma$, we obtain an automaton of size $O(n \sigma)$ and delay $O(1)$, thus matching the bound for the standard subsequence automaton construction. Finally, we generalize the result to multiple strings. The key component of our result is a novel hierarchical automata construction of independent interest. 
\end{abstract}


\section{Introduction}

Let $S$ be a string of length $n$ with characters from an alphabet of size $\sigma$. A \emph{subsequence} of $S$ is any string obtained by deleting zero or more characters from $S$. The \emph{subsequence automaton} (often called the \emph{directed acyclic subsequence graph}) is the minimal deterministic finite automaton accepting all subsequences of $S$. Baeza-Yates~\cite{BaezaYates1991} initiated the study of subsequence automata. They presented a simple construction using $O(n\sigma)$ size (size denotes the total number of states \emph{and} transitions) and showed that this bound is optimal in the sense that there are subsequence automata of size at least $\Omega(n \sigma)$. They also considered variations with encoded input strings and multiple strings. Subsequently, several researchers have further studied subsequence automata (and its variants)~\cite{tronivcek2005size,CMT2003,crochemore1999directed,Tronivcek2001episode,hoshino2000online,farhana2010finite,bannai2003inferring,tronivcek1999operations}. See also the surveys by Tron\'{\i}\v{c}ek~\cite{Tronicek2001,tronicek2003common}. The general problem of \emph{subsequence indexing}, not limited to automata based solutions, is investigated by Bille et al.~\cite{BFC2008}.

In this paper, we consider subsequence automata in the context of \emph{default transitions}, that is, special transitions to be taken only if none of the regular transitions match the current character, and which do not consume the current character. Each state has at most one default transition and hence the automaton remains deterministic. The key point of default transitions is to reduce the size of standard automata at the cost of introducing a \emph{delay}, i.e., the maximum number of consecutive default transition followed before consuming a character. 
For instance, given a pattern string of length $m$ the classic Knuth-Morris-Pratt (KMP)~\cite{KMP1977} string matching algorithm may be viewed as an automaton with default transitions (typically referred to as \emph{failure transitions}). This automaton has size $O(m)$, whereas the standard automaton with no default transitions would need $\Theta(m\sigma)$ space.
The delay of the automaton in the KMP algorithm is either $O(m)$ or $O(\log m)$ depending on the version. 
Similarly, the Aho-Corasick string matching algorithm for multiple strings may also be viewed as an automaton with default transitions~\cite{AC1975}. More recently, default transitions have also been used extensively to significantly reduce sizes of deterministic automata for regular expression~\cite{kumar2006algorithms,hayes2007dpico}. The main idea is to effectively enable states with large overlapping identical sets of outgoing transitions to "share" outgoing transitions using default transitions. 

Surprisingly, no non-trivial bounds for subsequence automata with default transitions are known. Naively\label{lab:naive-single}, we can immediately obtain an $O(n\sigma)$ size solution with $O(1)$ delay by using the standard subsequence automaton (without default transitions). At the other extreme, we can build an automaton with $n+1$ states (each corresponding to a prefix of $S$) with a standard and a default transition from the state corresponding to the $i$th prefix to the state corresponding to the $i+1$st prefix (the standard transition is labeled $S[i+1]$). It is straightforward to show that this leads to an $O(n)$ size solution with $O(n)$ delay. Our main result is a substantially improved trade-off between the size and delay of the subsequence automaton:

\begin{theorem}
\label{thm:main}
Let $S$ be a string of $n$ characters from an alphabet of size $\sigma$. For any integer parameter $k$, $1 < k \leq \sigma$, we can construct a subsequence automaton with default transitions of size $O(nk\log_k\sigma)$ and delay $O(\log_k \sigma)$.
\end{theorem}

Hence, with $k = 2$ we obtain an automaton of size $O(n \log \sigma)$ and delay $O(\log \sigma)$. On the other extreme, with $k = \sigma$, we obtain an automaton of size $O(n \sigma)$ and delay $O(1)$, thus matching the bound for the standard subsequence automaton construction. 

To obtain our result, we first introduce the \emph{level automaton}. Intuitively, this automaton uses the same states as the standard solution, but hierarchically orders them in a tree-like structure and samples a selection of their original transitions based on their position in the tree, and adds a default transition to the next state on a higher level. We show how to do this efficiently leading to a solution with $O(n \log n)$ size and $O(\log n)$ delay. To achieve our full trade-off from Theorem~\ref{thm:main} we show how to augment the construction with additional ideas for small alphabets and generalize the level automaton with parameter $k$, $1 < k \leq \sigma$, where large $k$ reduces the height of the tree but increases the number of transitions. In the final section we generalize the result to \emph{multiple} strings.

\section{Preliminaries}
A \emph{deterministic finite automaton} (DFA) is a tuple $A = (Q, \Sigma, \delta,  q_0, F)$ where $Q$ is a set of nodes called \emph{states}, $\delta$ is a set of labeled directed edges between states, called \emph{transitions}, where each label is a character from the alphabet $\Sigma$, $q_0\in Q$ is the \emph{initial} state and $F\subseteq Q$ is a set of \emph{accepting} states. No two outgoing transitions from the same state have the same label. The DFA is \emph{incomplete} in the sense that every state does not contain transitions for every character in $\Sigma$.
 The \emph{size} of $A$ is the sum of the number of states and transitions.

We can think of $A$ as an \emph{edge-labeled directed graph}.
Given a string $P$ and a path $p$ in $A$ we say that $p$ and $P$ match if the concatenation of the labels on the transitions in $p$ is $P$. We say that $A$ \emph{accepts} a string $P$ if there is a path in $A$, from $q_0$ to any state in $F$, that matches $P$. Otherwise $A$ \emph{rejects} $P$.


A \emph{deterministic finite automaton with default transitions} is a deterministic finite automaton $AD$ where each state can have a single unlabeled default transition. Given a string $P$ and a path $p$ in $AD$ we define a match between $P$ and $p$ as before, with the exception that for any default transition $d$ in $p$ the corresponding character in $P$ cannot match any standard transition out of the start state of $d$. Definition of accepted and rejected strings are as before. The \emph{delay} of $AD$ is the maximum length of any path matching a single character, i.e., if the delay of $AD$ is $d$ then we follow at most $d-1$ default transitions for every character that is matched in $P$.

\begin{figure}
\centering
\begin{tikzpicture}
  \draw (-1*\constantx,2*\constanty) node(s0) [circle, draw, scale=\nodescale] {}
        (0*\constantx,2*\constanty) node(s1) [circle, draw, scale=\nodescale] {}
        (1*\constantx,2*\constanty) node(s2) [circle, draw,scale=\nodescale] {}
        (2*\constantx,2*\constanty) node(s3) [circle, draw,scale=\nodescale] {}
        (3*\constantx,2*\constanty) node(s4) [circle, draw,scale=\nodescale] {}
        (4*\constantx,2*\constanty) node(s5) [circle, draw,scale=\nodescale] {}
        (5*\constantx,2*\constanty) node(s6) [circle, draw,scale=\nodescale] {};
  \draw (0*\constantx,0.5*\constanty) node(a1) [scale=\letterscale] {$a$}
        (1*\constantx,0.5*\constanty) node(b2) [scale=\letterscale] {$b$}
        (2*\constantx,0.5*\constanty) node(a3) [scale=\letterscale] {$a$}
        (3*\constantx,0.5*\constanty) node(d4) [scale=\letterscale] {$d$}
        (4*\constantx,0.5*\constanty) node(c5) [scale=\letterscale] {$c$}
        (5*\constantx,0.5*\constanty) node(a6) [scale=\letterscale] {$a$};
  \draw[->] (s0) edge node[above, scale=\letterfigsc] {$a$} (s1);
  \draw[->] (s0) edge[bend right] node[below, scale=\letterfigsc] {$b$} (s2);
  \draw[->] (s0) edge[bend left=60] node[above, scale=\letterfigsc] {$c$} (s5);
  \draw[->] (s0) edge[bend right=50] node[above, scale=\letterfigsc] {$d$} (s4);
  \draw[->] (s1) edge[bend left=60] node[above,scale=\letterfigsc] {$a$} (s3);
  \draw[->] (s1) edge node[above,scale=\letterfigsc] {$b$} (s2);
  \draw[->] (s1) edge[bend right] node[above,scale=\letterfigsc] {$c$} (s5);
  \draw[->] (s1) edge[bend left=90] node[above,scale=\letterfigsc] {$d$} (s4);
  \draw[->] (s2) edge[] node[above,scale=\letterfigsc] {$a$} (s3);
  \draw[->] (s2) edge[bend right] node[above,scale=\letterfigsc] {$c$} (s5);
  \draw[->] (s2) edge[bend left=80] node[above,scale=\letterfigsc] {$d$} (s4);
  \draw[->] (s3) edge[bend left=50] node[above,scale=\letterfigsc] {$a$} (s6);
  \draw[->] (s3) edge[bend right] node[above,scale=\letterfigsc] {$c$} (s5);
  \draw[->] (s3) edge[] node[above,scale=\letterfigsc] {$d$} (s4);
  \draw[->] (s4) edge[bend left] node[above,scale=\letterfigsc] {$a$} (s6);
  \draw[->] (s4) edge[] node[above,scale=\letterfigsc] {$c$} (s5);
  \draw[->] (s5) edge[] node[above,scale=\letterfigsc] {$a$} (s6);
    

  \end{tikzpicture}
\caption{\label{fig:simpledasg} An example of an SA constructed from the string $abadca$.}
\end{figure}
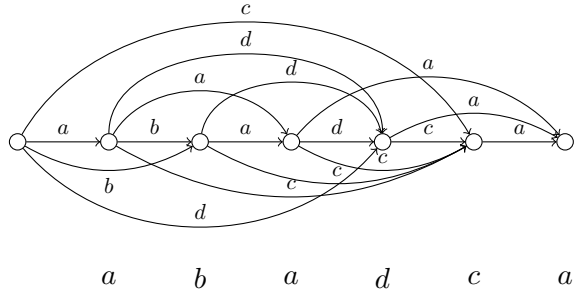

A \emph{subsequence} of $S$ is a string $P$, obtained by removing zero or more occurrences of characters from $S$. The alphabet of the string $S$ is denoted by $\Sigma(S)$.
A \emph{subsequence automaton} constructed from $S$, is a deterministic finite automaton that accepts string $P$ iff $P$ is a subsequence of $S$. A subsequence automaton construction is presented in~\cite{BaezaYates1991}. This construction is often called the \emph{directed acyclic subsequence graph} or DASG, but here we denote it SA.
The SA has $n+1$ states, all accepting, that we identify with the integers $\{0,1, \ldots, n\}$.
For each state $s$, $0 \leq s \leq n$, we have the following transitions:
\begin{itemize}
\item For each character $\alpha$ in $\Sigma(S[s + 1, n])$, there is a transition labeled $\alpha$ to the smallest state $s' > s$ such that $S[s']= \alpha$.
\end{itemize}
The SA has size $O(n\sigma)$ since every state can have at most $\sigma$ transitions.
An example of an SA is given in Figure~\ref{fig:simpledasg}.

A \emph{subsequence automaton with default transitions} constructed from $S$, denoted SAD, is a deterministic finite automaton with default transitions that accepts string $P$ iff $P$ is a subsequence of $S$. 

The next section explores different configurations of transitions and default transitions in SADs.

\section{New Trade-Offs for Subsequence Automata.}
We now present a new trade-off for subsequence automata, with default transitions. We will gradually refine our construction until we obtain an automaton that gives the result presented in Theorem~\ref{thm:main}. In each construction we have $n+1$ states that we identify with the integers $\{0,1, \ldots, n\}$. Each of these states represents a prefix of the string $S$ and are all accepting states.
We first present the \emph{level automaton} that gives the first non-trivial trade-off that exploits default transitions. 
The general idea is to construct a hierarchy of states, such that every path that only uses default transitions is guaranteed to go through states where the outdegree increases at least exponentially.
The level automaton is a SAD of size $O(n\log n)$ and delay $O(\log n)$. 
By arguing that any path going through a state with outdegree $\sigma$ will do so by taking a regular transition, we are able to improve both the size and delay of the level automaton. 
This results in the \emph{alphabet-aware level automaton} which is a SAD of size $O(n\log \sigma)$ and delay $O(\log \sigma)$. Finally we present a generalized  construction that gives a trade-off between size and delay by letting parameter $k$, $1 < k \leq \sigma$, be the base of the exponential increase in outdegree on paths with only default transitions. This SAD has size $O(nk\log_k \sigma)$ and delay $O(\log_k \sigma)$. With $k=2$ we get an automaton of size $O(n\log \sigma)$ and delay $O(\log \sigma)$. In the other extreme, for $k=\sigma$ we get an automaton of size $O(n\sigma)$ and delay $O(1)$.
\subsection{Level Automaton}
The level automaton is a SAD with $n+1$ states that we identify with the integers $\{0,1, \ldots, n\}$. All states are accepting. For each state $i>0$, we associate a level, $\level(i)$, given by:
\begin{gather*}
\level(i) =\max(\{x \;|\; i \bmod 2^x = 0 \})
\end{gather*}

Hence, $\level(i)$ is the exponent of the largest power of two that divides $i$. The level function is in the literature known as the ruler function. 
We do not associate any level with state 0. Note that the maximum level of any state is $\log_2 n$.
For a nonnegative integer $s$, we define $\overline{s}$ to be the smallest integer $\overline{s} > s$ such that $\level(\overline{s}) \geq \level(s) +1$. 

\begin{figure}
\centering
\begin{tikzpicture}
  \draw (-1*\constantx,2*\constanty) node(s0) [circle, draw, scale=\nodescale] {}
        (0*\constantx,2*\constanty) node(s1) [circle, draw, scale=\nodescale] {}
        (1*\constantx,3*\constanty) node(s2) [circle, draw,scale=\nodescale] {}
        (2*\constantx,2*\constanty) node(s3) [circle, draw,scale=\nodescale] {}
        (3*\constantx,4*\constanty) node(s4) [circle, draw,scale=\nodescale] {}
        (4*\constantx,2*\constanty) node(s5) [circle, draw,scale=\nodescale] {}
        (5*\constantx,3*\constanty) node(s6) [circle, draw,scale=\nodescale] {}
        (6*\constantx,2*\constanty) node(s7) [circle, draw, scale=\nodescale] {}
        (7*\constantx,5*\constanty) node(s8) [circle, draw,scale=\nodescale] {}
        (8*\constantx,2*\constanty) node(s9) [circle, draw,scale=\nodescale] {}
        (9*\constantx,3*\constanty) node(s10) [circle, draw,scale=\nodescale] {}
        (10*\constantx,2*\constanty) node(s11) [circle, draw,scale=\nodescale] {}
        (11*\constantx,4*\constanty) node(s12) [circle, draw,scale=\nodescale] {};
  \draw (0*\constantx,1*\constanty) node(a1) [scale=\letterscale] {$a$}
        (1*\constantx,1*\constanty) node(b2) [scale=\letterscale] {$b$}
        (2*\constantx,1*\constanty) node(a3) [scale=\letterscale] {$a$}
        (3*\constantx,1*\constanty) node(d4) [scale=\letterscale] {$c$}
        (4*\constantx,1*\constanty) node(c5) [scale=\letterscale] {$b$}
        (5*\constantx,1*\constanty) node(a6) [scale=\letterscale] {$a$}
        (6*\constantx,1*\constanty) node(b2) [scale=\letterscale] {$b$}
        (7*\constantx,1*\constanty) node(a3) [scale=\letterscale] {$c$}
        (8*\constantx,1*\constanty) node(d4) [scale=\letterscale] {$a$}
        (9*\constantx,1*\constanty) node(c5) [scale=\letterscale] {$b$}
        (10*\constantx,1*\constanty) node(a6) [scale=\letterscale] {$a$}
        (11*\constantx,1*\constanty) node(c5) [scale=\letterscale] {$d$};
        
  \draw[->] (s0) edge node[above, scale=\letterfigsc] {$a$} (s1);
  \draw[->] (s0) edge[bend right=15,style=dashed] node[below, scale=\letterfigsc] {} (s1);
  \draw[->] (s1) edge[] node[right, scale=\letterfigsc] {$b$} (s2);
  \draw[->] (s1) edge[bend left=15,style=dashed] node[above, scale=\letterfigsc] {} (s2);
  \draw[->] (s2) edge[] node[right,scale=\letterfigsc] {$a$} (s3);
  \draw[->] (s2) edge node[above,scale=\letterfigsc] {$c$} (s4);
  \draw[->] (s2) edge[bend left,style=dashed] node[above,scale=\letterfigsc] {} (s4);
  
  \draw[->] (s3) edge[] node[right,scale=\letterfigsc] {$c$} (s4);
  \draw[->] (s3) edge[bend left=15,style=dashed] node[above,scale=\letterfigsc] {} (s4);

  \draw[->] (s4) edge[] node[right,scale=\letterfigsc] {$b$} (s5);
  \draw[->] (s4) edge[] node[above,scale=\letterfigsc] {$a$} (s6);
  \draw[->] (s4) edge[] node[above,scale=\letterfigsc] {$c$} (s8);
  \draw[->] (s4) edge[style=dashed,bend left=15] node[above,scale=\letterfigsc] {} (s8);

  \draw[->] (s5) edge[] node[below,scale=\letterfigsc] {$a$} (s6);
  \draw[->] (s5) edge[bend left=15, style=dashed] node[above,scale=\letterfigsc] {} (s6);

  \draw[->] (s6) edge[] node[above,scale=\letterfigsc] {$b$} (s7);
  \draw[->] (s6) edge[] node[left,scale=\letterfigsc] {$c$} (s8);
  \draw[->] (s6) edge[style=dashed,bend left=15] node[above,scale=\letterfigsc] {} (s8);
  
  \draw[->] (s7) edge[] node[right,scale=\letterfigsc] {$c$} (s8);
  \draw[->] (s7) edge[style=dashed,bend left=15] node[above,scale=\letterfigsc] {} (s8);

  \draw[->] (s8) edge[] node[right,scale=\letterfigsc] {$a$} (s9);
  \draw[->] (s8) edge[] node[right,scale=\letterfigsc] {$b$} (s10);
  \draw[->] (s8) edge[] node[above,scale=\letterfigsc] {$d$} (s12);

  \draw[->] (s9) edge[] node[right,scale=\letterfigsc] {$b$} (s10);
  \draw[->] (s9) edge[bend left=15,style=dashed] node[above,scale=\letterfigsc] {} (s10);

  \draw[->] (s10) edge[] node[above,scale=\letterfigsc] {$a$} (s11);
  \draw[->] (s10) edge[] node[below,scale=\letterfigsc] {$d$} (s12);
  \draw[->] (s10) edge[bend left=15,style=dashed] node[above,scale=\letterfigsc] {} (s12);

  \draw[->] (s11) edge[] node[right,scale=\letterfigsc] {$d$} (s12);
  \draw[->] (s11) edge[bend left=15,style=dashed] node[above,scale=\letterfigsc] {} (s12);

    

  \end{tikzpicture}
\caption{\label{fig:levelautomaton}The level automaton constructed from the string $abacbabcabad$.}
\end{figure}
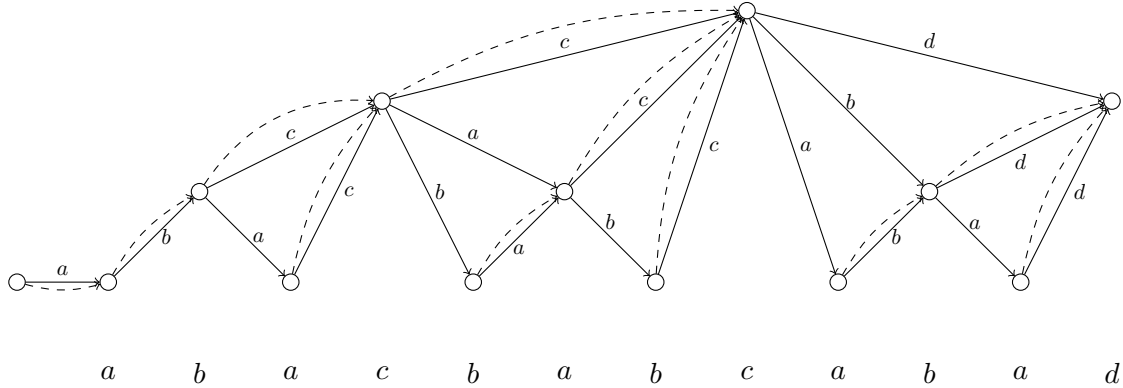

The transitions in the level automaton are as follows:
From state 0 we have a default transition to state 1 and a regular transition to state 1 with label $S[1]$. For every other state $s$, $1 \leq s \leq n$, we have the following transitions:
\begin{itemize}
\item A default transition to state $\overline{s}$. If no such state exist, the state $s$ does not have a default transition.
\item For each character $\alpha$ in $\Sigma(S[s + 1, \min(\overline{s}, n)])$, there is a transition labeled $\alpha$ to the smallest state $s' > s$ such that $S[s']= \alpha$.
\end{itemize}
An example of the level automaton constructed from the string $abacbabcabad$ and alphabet $\{a,b,c,d\}$ is given in Figure~\ref{fig:levelautomaton}. The dashed arrows denote default transitions and the vertical position of the states denotes their level.



We first show that the level automaton is a SAD for $S$, i.e., the level automaton accepts a string iff the string is a subsequence of $S$. To do so suppose that $P$ is a string of length $m$ accepted by the level automaton and let $s_1,s_2, \ldots , s_m$ be the sequence of states visited with regular transitions on the path that accepts $P$. From the definition of the transition function, we know that if a transition with label $\alpha$ leads to state $s'$, then $S[s']=\alpha$. This means that $S[s_1]S[s_2]\ldots S[s_m]$ spells out a subsequence of $S$ if the sequence $s_1, s_2, \ldots, s_m$ is strictly increasing. From the definition of the transitions, a state $s$ only have transitions to states $s'$ if $s'> s$. Hence, the sequence is strictly increasing.

For the other direction, we show that the level automaton simulates the SA. At each state $s$, trying to match character $\alpha$, we find the smallest state $s'>s$ such that $s'$ has an incoming transition with label $\alpha$: By the construction, either $s$ has an outgoing transition leading directly to $s'$ \emph{or} we follow default transitions until reaching the first state with a transition to $s'$. This means that the states visited with standard transitions in the level automaton are the same states that are visited in the SA. Since the SA accepts all subsequences of $S$ this must also hold for the level automaton.

\subsubsection{Analysis}
The following shows that the number of outgoing transitions increase with a factor two when the level increase by one.
For all $s>0$, we have the following property of $\overline{s}$ and $\level(s)$:
\begin{gather}
\overline{s} - s = 2^{\level(s)}\label{sec:overlinesminuss}
\end{gather}
By definition, $2^{\level(s)}$ divides $s$. This means that we can write $s$ as  $c\cdot 2^{\level(s)}$, where $c$ is a uneven positive integer. We know that $c$ is uneven because $2^{\level(s)}$ is the largest power of two that divides $s$. The next integer, larger than $s$, that $2^{\level(s)}$ divides is $s'= s + 2^{\level(s)}$. This means that $\overline{s}\geq s'$. We can rewrite $s'$ as follows: $s'= s + 2^{\level(s)}=c\cdot 2^{\level(s)} + 2^{\level(s)}=(c+1)\cdot 2^{\level(s)}$. Since $c$ is uneven we know that $c+1$ is even so we can rewrite $s'$ further: $s'=\frac{(c+1)}{2}\cdot 2^{\level(s)+1}$. This shows that $2^{\level(s)+1}$ divides $s'$ which means that $s'=\overline{s}$ and we conclude that $\overline{s} - s = 2^{\level(s)}$.

Since the maximal level of any state is $\log_2 n$ and the level increase every time we follow a default transition, the delay of the level automaton is $O(\log n)$.



At each level $l$ we have $O(n/2^{l+1})$ states, since every $2^l$th state is divided by $2^l$, and $2^l$ is the largest divisor in every second of these cases. Since $\overline{s} - s = 2^{\level(s)}$ each state at level $l$ has at most $2^{l}+1$ outgoing transitions. Therefore, each level contribute with size at most $n/2^{l+1} \cdot (2^{l}+1)= O(n)$. Since we have at most $O(\log n)$ levels, the total size becomes $O(n\log n)$.

In summary, we have shown the following result.
\begin{lemma}\label{lemma:levelaut}
Let $S$ be a string of $n$ characters. We can construct a subsequence automaton with default transitions of size $O(n\log n)$ and delay $O(\log n)$.
\end{lemma}
\subsection{Alphabet-aware level automaton}
We introduce the \emph{Alphabet-aware level automaton}. When the level automaton reaches a state $s$ where $\overline{s}-s\geq\sigma$, then $s$ can have up to $\sigma$ outgoing transitions without violating the space analysis above. The level automaton only has a transition for each character in $\Sigma(S[s+1, \min(\overline{s}, n)])$. Hence, for all states $s$ in the alphabet-aware level automaton where $\overline{s} -s \geq \sigma$, we let $s$ have a transition for each symbol $\alpha$ in $\Sigma$, to the smallest state $s'>s$ such that $S[s']=\alpha$. No matching path can take a default transition from a state with $\sigma$ outgoing transitions. Hence, states with $\sigma$ outgoing transitions do not need default transitions.

We change the level function to reflect this. For each state $1\leq i \leq n$ we have that:
\begin{gather*}
\level(i)= \min(\lceil \log_2 \sigma \rceil, \max(\{x \;|\; i \bmod 2^x = 0 \}))
\end{gather*}

The transitions in the alphabet-aware level automaton is as follows:
From state 0 we have a default transition to state 1 and a regular transition to state 1 with label $S[1]$.
For every other state $s$, $1 \leq s \leq n$, we have the following transitions:
\begin{itemize}
\item A default transition to state $\overline{s}$. If no such state exist, the state $s$ does not have a default transition.
\item If $\overline{s} - s < \sigma$ then for each character $\alpha$ in $\Sigma(S[s + 1, \min(\overline{s}, n)])$, there is a transition labeled $\alpha$ to the smallest state $s' > s$ such that $S[s']= \alpha$.
\item If $\overline{s} - s \geq \sigma$ then for each character $\alpha$ in $\Sigma(S[s + 1,  n])$, there is a transition labeled $\alpha$ to the smallest state $s' > s$ such that $S[s']= \alpha$.
\end{itemize}

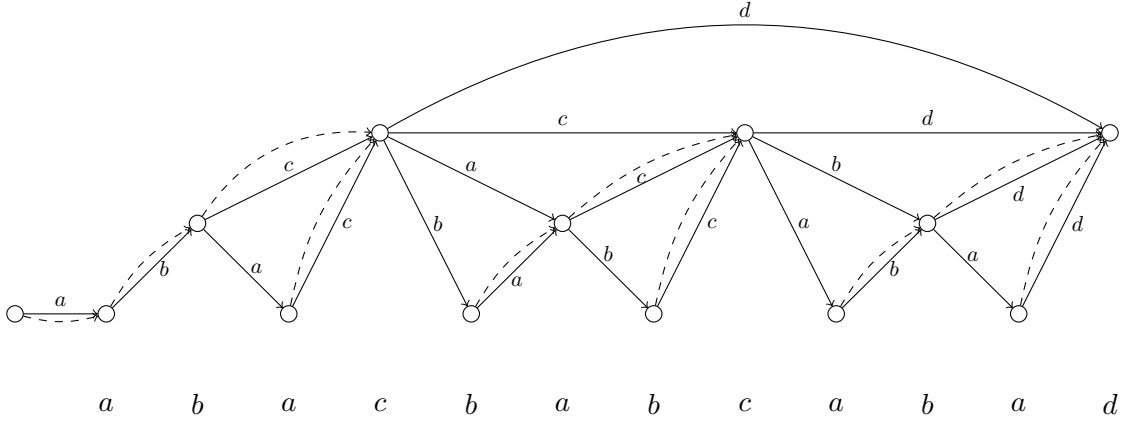
\begin{figure}
\centering
\begin{tikzpicture}
  \draw (-1*\constantx,2*\constanty) node(s0) [circle, draw, scale=\nodescale] {}
        (0*\constantx,2*\constanty) node(s1) [circle, draw, scale=\nodescale] {}
        (1*\constantx,3*\constanty) node(s2) [circle, draw,scale=\nodescale] {}
        (2*\constantx,2*\constanty) node(s3) [circle, draw,scale=\nodescale] {}
        (3*\constantx,4*\constanty) node(s4) [circle, draw,scale=\nodescale] {}
        (4*\constantx,2*\constanty) node(s5) [circle, draw,scale=\nodescale] {}
        (5*\constantx,3*\constanty) node(s6) [circle, draw,scale=\nodescale] {}
        (6*\constantx,2*\constanty) node(s7) [circle, draw, scale=\nodescale] {}
        (7*\constantx,4*\constanty) node(s8) [circle, draw,scale=\nodescale] {}
        (8*\constantx,2*\constanty) node(s9) [circle, draw,scale=\nodescale] {}
        (9*\constantx,3*\constanty) node(s10) [circle, draw,scale=\nodescale] {}
        (10*\constantx,2*\constanty) node(s11) [circle, draw,scale=\nodescale] {}
        (11*\constantx,4*\constanty) node(s12) [circle, draw,scale=\nodescale] {};
  \draw (0*\constantx,1*\constanty) node(a1) [scale=\letterscale] {$a$}
        (1*\constantx,1*\constanty) node(b2) [scale=\letterscale] {$b$}
        (2*\constantx,1*\constanty) node(a3) [scale=\letterscale] {$a$}
        (3*\constantx,1*\constanty) node(d4) [scale=\letterscale] {$c$}
        (4*\constantx,1*\constanty) node(c5) [scale=\letterscale] {$b$}
        (5*\constantx,1*\constanty) node(a6) [scale=\letterscale] {$a$}
        (6*\constantx,1*\constanty) node(b2) [scale=\letterscale] {$b$}
        (7*\constantx,1*\constanty) node(a3) [scale=\letterscale] {$c$}
        (8*\constantx,1*\constanty) node(d4) [scale=\letterscale] {$a$}
        (9*\constantx,1*\constanty) node(c5) [scale=\letterscale] {$b$}
        (10*\constantx,1*\constanty) node(a6) [scale=\letterscale] {$a$}
        (11*\constantx,1*\constanty) node(c5) [scale=\letterscale] {$d$};
        
  \draw[->] (s0) edge node[above, scale=\letterfigsc] {$a$} (s1);
  \draw[->] (s0) edge[bend right=15,style=dashed] node[below, scale=\letterfigsc] {} (s1);
  \draw[->] (s1) edge[] node[right, scale=\letterfigsc] {$b$} (s2);
  \draw[->] (s1) edge[bend left=15,style=dashed] node[above, scale=\letterfigsc] {} (s2);
  \draw[->] (s2) edge[] node[right,scale=\letterfigsc] {$a$} (s3);
  \draw[->] (s2) edge node[above,scale=\letterfigsc] {$c$} (s4);
  \draw[->] (s2) edge[bend left,style=dashed] node[above,scale=\letterfigsc] {} (s4);
  
  \draw[->] (s3) edge[] node[right,scale=\letterfigsc] {$c$} (s4);
  \draw[->] (s3) edge[bend left=15,style=dashed] node[above,scale=\letterfigsc] {} (s4);

  \draw[->] (s4) edge[] node[right,scale=\letterfigsc] {$b$} (s5);
  \draw[->] (s4) edge[] node[above,scale=\letterfigsc] {$a$} (s6);
  \draw[->] (s4) edge[] node[above,scale=\letterfigsc] {$c$} (s8);
  \draw[->] (s4) edge[bend left] node[above,scale=\letterfigsc] {$d$} (s12);

  \draw[->] (s5) edge[] node[below,scale=\letterfigsc] {$a$} (s6);
  \draw[->] (s5) edge[bend left=15, style=dashed] node[above,scale=\letterfigsc] {} (s6);

  \draw[->] (s6) edge[] node[above,scale=\letterfigsc] {$b$} (s7);
  \draw[->] (s6) edge[] node[left,scale=\letterfigsc] {$c$} (s8);
  \draw[->] (s6) edge[style=dashed,bend left=15] node[above,scale=\letterfigsc] {} (s8);
  
  \draw[->] (s7) edge[] node[right,scale=\letterfigsc] {$c$} (s8);
  \draw[->] (s7) edge[style=dashed,bend left=15] node[above,scale=\letterfigsc] {} (s8);

  \draw[->] (s8) edge[] node[right,scale=\letterfigsc] {$a$} (s9);
  \draw[->] (s8) edge[] node[above,scale=\letterfigsc] {$b$} (s10);
  \draw[->] (s8) edge[] node[above,scale=\letterfigsc] {$d$} (s12);

  \draw[->] (s9) edge[] node[right,scale=\letterfigsc] {$b$} (s10);
  \draw[->] (s9) edge[bend left=15,style=dashed] node[above,scale=\letterfigsc] {} (s10);

  \draw[->] (s10) edge[] node[above,scale=\letterfigsc] {$a$} (s11);
  \draw[->] (s10) edge[] node[below,scale=\letterfigsc] {$d$} (s12);
  \draw[->] (s10) edge[bend left=15,style=dashed] node[above,scale=\letterfigsc] {} (s12);

  \draw[->] (s11) edge[] node[right,scale=\letterfigsc] {$d$} (s12);
  \draw[->] (s11) edge[bend left=15,style=dashed] node[above,scale=\letterfigsc] {} (s12);

    

  \end{tikzpicture}
\caption{\label{fig:alphalevelautomaton}The alphabet level automaton constructed from the string $abacbabcabad$.}
\end{figure}

An example of the alphabet-aware level automaton constructed from the string $abacbabcabad$ and alphabet $\{a,b,c,d\}$ is given in Figure~\ref{fig:alphalevelautomaton}. The level automaton in Figure~\ref{fig:levelautomaton} is constructed from the same string and the same alphabet. For comparison, state 4 in Figure~\ref{fig:alphalevelautomaton} now has outdegree $\sigma$ and has transitions to the first succeeding occurrence of any unique character and state 8 has been constrained to level $\lceil \log_2 \sigma \rceil$.

The alphabet-aware level automaton is a SAD by the same arguments that led to Lemma~\ref{lemma:levelaut}.


The delay is now bounded by $O(\log \sigma)$ since no state is assigned a level higher than $\lceil \log_2 \sigma \rceil$.
The size of each level is still $O(n)$. Hence, the total size becomes $O(n\log \sigma)$

In summary, we have shown the following result.
\begin{lemma}\label{lemma:alphalevel}
  Let $S$ be a string of $n$ characters. We can construct a SAD of $S$ with size $O(n\log \sigma)$ and delay $O(\log \sigma)$.
\end{lemma}

\subsection{Full trade-off}
We can generalize the construction above by introducing parameter $k$, $1<k\leq\sigma$, which is the base of the exponential increase in outdegree of states on every path that only uses default transitions. Now, when we follow a default transition from $s$ to $\overline{s}$, the number of outgoing transitions increase with a factor $k$ instead of a factor 2.
This gives a trade-off between size and delay in the SAD determined by $k$. Increasing $k$ gives a shorter delay of the SAD but increases the size and vice versa.

Each state, except state 0, is still associated with a level, but we need to generalize the level function to account for the parameter $k$.
For every $k$ and $i$ we have that:
\begin{gather*}
\level(i, k)= \min(\lceil \log_k \sigma \rceil, \max(\{x \;|\; i \bmod k^x = 0 \}))
\end{gather*}

Now, the level function gives the largest power of $k$ that divides $i$. 

The transitions in the generalized alphabet-aware level automaton is as follows:
From state 0 we have a default transition to state 1 and a regular transition to state 1 with label $S[1]$.
For every other state $s$, $1 \leq s \leq n$, we have the following transitions:
\begin{itemize}
\item A default transition to state $\overline{s}$. If no such state exist, the state $s$ does not have a default transition.
\item If $\overline{s} - s < \sigma$ then for each character $\alpha$ in $\Sigma(S[s + 1, \min(\overline{s}, n)])$, there is a transition labeled $\alpha$ to the smallest state $s' > s$ such that $S[s']= \alpha$.
\item If $\overline{s} - s \geq \sigma$ then for each character $\alpha$ in $\Sigma(S[s + 1,  n])$, there is a transition labeled $\alpha$ to the smallest state $s' > s$ such that $S[s']= \alpha$.
\end{itemize}

We can show that the generalized alphabet-aware level automaton is a SAD by the same arguments that led to Lemma~\ref{lemma:alphalevel}.
\subsubsection{Analysis}
The delay is bounded by $O(\log_k \sigma)$ because no state is assigned a level higher than $\lceil \log_k \sigma \rceil$.

With the new definition of the level function we have that 
\begin{gather*}
\overline{s} - s \leq k^{\level(s,k)+1}
\end{gather*}
for all $s>0$.
This expression bounds the number of outgoing transitions from state $s$.

At level $l$ we have $O(n(k-1)/(k^{l+1}))$ states each with $O(k^{l+1})$ outgoing transitions such that each level has size $O(nk)$. The size of the automaton becomes $O(nk\log_k \sigma)$ because we have $O(\log_k \sigma)$ levels.

In summary, we have shown Theorem~\ref{thm:main}.

\section{Subsequence automata for multiple strings}



Tron\'{\i}\v{c}ek et al.~\cite{CMT2003} generalizes the simple subsequence automaton to multiple strings. Given a set of strings $\mathcal{S}=\{S_1, S_2,\ldots. S_N\}$ of length $n1,n2, \ldots, n_N$, two types of automata are presented: The subsequence automaton accepts a pattern $P$ iff $P$ is a subsequence of \emph{some} string in $\mathcal{S}$ and the \emph{common} subsequence automaton accepts $P$ iff $P$ is a subsequence of \emph{every} string in $\mathcal{S}$. When $\mathcal{S}=\{S_1, S_2,\ldots, S_n\}$ and $ n_1 = n_2 =\ldots= n_N =n$, a $\Omega(n^N/(N+1)^NN!))$ lower bound on the number of states is shown for the subsequence automaton~\cite{tronivcek2005size}. For both automata, the number of states is trivially upper bounded by $O( n_1 \cdot  n_2  \cdot \ldots \cdot  n_N )$ such that the total size becomes $O(\sigma\cdot n_1 \cdot  n_2  \cdot \ldots \cdot  n_N )$. We can reduce the size of these automata by augmenting them with default transitions. This generalization is in the same spirit as the naive generalization of the single string automaton in section \ref{lab:naive-single}: We introduce default transitions and save a factor $\sigma$ in the space but also introduce a delay. Consider the naive common subsequence automaton with default transitions: For two strings $S_1,S_2$ we have $ n_1 \cdot  n_2 +1$ states that we identify with the points $\{1,\ldots,  n_1 \}\times\{1,\ldots,  n_2 \}\cup \{(0,0)\}$. For each state $(s_1,s_2)$ we have the following transitions:

\begin{itemize}
\item A default transition to state $(s_1+1,s_2+1)$. If no such state exist, the state $(s_1,s_2)$ does not have a default transition.
\item If character $S_1[s_1+1]$ is in $\Sigma(S_2[s_2+1,  n_2 ])$, there is a transition labeled $S_1[s_1+1]$ to the state $(s_1+1,s_2')$ such that $s_2'>s_2$ is the minimal index where $S_2[s_2']= S_1[s_1+1]$.
\item If character $S_2[s_2+1]$ is in $\Sigma(S_1[s_1+1,  n_1 ])$, there is a transition labeled $S_2[s_2+1]$ to the state $(s_1',s_2+1)$ such that $s_1'>s_1$ is the minimal index where $S_1[s_1']= S_2[s_2+1]$.
\end{itemize}

The states of the automaton represents the progression in $S_1$ and $S_2$, such that state $(s_1, s_2)$ represents that subsequences of the prefixes $S_1[1,s_1]$ and $S_2[1,s_2]$ have been used to match a prefix of $P$. Each state $(s_1, s_2)$ considers the symbols $S_1[s_1+1]$ and $S_2[s_2+1]$ for matching with the next symbol in $P$. If this is not possible, a default transition is followed to state $(s_1+1, s_2+1)$.

For this automaton the size is $O( n_1 \cdot  n_2 )$ and the delay is $O(\min( n_1 , n_2 ))$. Hence, we save a $\sigma$ factor in the size, but introduce a significant delay. As we did for the subsequence automaton for a single string, we introduce a level automaton that associates a level with each state. In this way we are able to reduce the delay significantly with only a small increase in size. For simplicity we only present our construction for the common subsequence automaton, but it follows immediately that the construction also applies to the subsequence automaton.

\subsection{The alphabet-aware level automaton for two strings}
The alphabet-aware level automaton for two strings $S_1,S_2$ have $ n_1 \cdot  n_2 +1$ states that we identify with the points $\{1,\ldots,  n_1 \}\times\{1,\ldots,  n_2 \}\cup \{(0,0)\}$. We define the \emph{diagonal} of a state $(i,j)$, as the set of states $\{(i+k, j+k) \;|\; 0 < i+k \leq  n_1 \textnormal{ and } 0 < j+k \leq  n_2 \}$. We say that states belong to the same diagonal if the diagonals of the states defines identical sets of states. For states $(s_1, s_2),(s_1',s_2')$ in the same diagonal, $(s_1, s_2)<(s_1',s_2')$ if $s_1 < s_1'$. For each state $(s_1, s_2)$ we associates the integer $\min(s_1,s_2)$, which is also its position in the diagonal, such that $(s_1, s_2)- (s_1', s_2')=\min(s_1,s_2)-\min(s_1',s_2')$ and $(s_1,s_2) \bmod x = \min(s_1,s_2) \bmod x$. With each diagonal of states, we associate a level structure identical to the one used in the alphabet-aware level automaton for a single string.  Now, when following a default transition from state $(s_1, s_2)$ to $(s_1+k, s_2+k)$, $k>0$, every unique symbol in $\Sigma(S_1[s_1+1, s_1+k])\cup \Sigma(S_2[s_2+1, s_2+k])$ contributes with a transition.
For each state $(i,j)$, $1\leq i \leq n_1,1\leq j \leq n_2$, we again associate a level:
\begin{gather*}
\level((i,j)) = \min(\left\lceil \log_2 \sigma \right\rceil, \max(\{x \;|\; (i,j) \bmod 2^x = 0 \}))
\end{gather*}


For a pair of positive integers $(i,j)$, we define $\overline{(i,j)}>(i,j)$ to be the smallest pair of integers in the same diagonal such that $\level((i,j)) < \level(\overline{(i,j)})$.

The alphabet-aware level automaton for two strings $S_1,S_2$ has the following transitions: State $(0,0)$ has a transition labeled $S_1[1]$ to state $(1, s_2)$ where $s_2$ is the minimal index such that $S_2[s_2]=S_1[1]$, a transition labeled $S_2[1]$ to state $(s_1, 1)$ where $s_1$ is the minimal index such that $S_1[s_1]=S_2[1]$ and a default transition to state $(1,1)$.  These transitions only exists if the indices exists. Every other state state $(s_1,s_2)$, where $(\overline{s_1}, \overline{s_2})= \overline{(s_1,s_2)}$, have the following transitions:
\begin{itemize}
\item A default transition to state $\overline{(s_1,s_2)}$. If no such state exist, the state $(s_1,s_2)$ does not have a default transition.
\item If $\overline{(s_1,s_2)} - (s_1,s_2) < \sigma$ then for each character $\alpha$ in $\Sigma(S_1[s_1 + 1, \min(\overline{s_1},  n_1 )]) \cup \Sigma(S_2[s_2 + 1, \min(\overline{s_2},  n_2 )])$, there is a transition labeled $\alpha$ to the state $(s_1',s_2')$, where $s_1'>s_1$ and $s_2'>s_2$ are the minimal indices such that $S_1[s_1']= S_2[s_2']= \alpha$.
\item If $\overline{(s_1,s_2)} - (s_1,s_2) \geq \sigma$ then for each character $\alpha$ in $\Sigma(S_1[s_1 + 1,  n_1 ]) \cup \Sigma(S_2[s_2 + 1,  n_2 ])$, there is a transition labeled $\alpha$ to the state $(s_1',s_2')$, where $s_1'>s_1$ and $s_2'>s_2$ are the minimal indices such that $S_1[s_1']=S_2[s_2']= \alpha$.
\end{itemize}

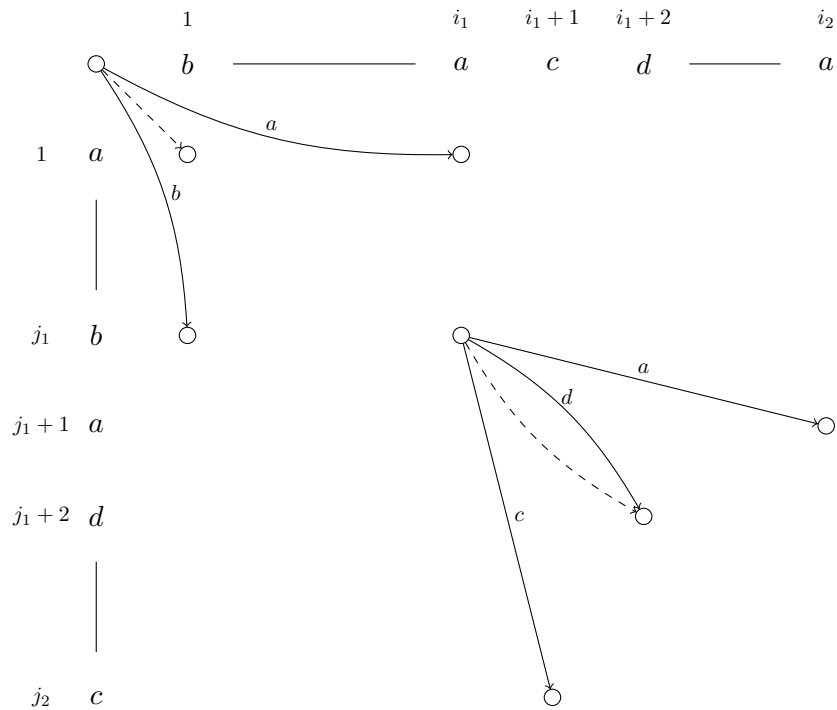
\begin{figure}
\centering
\begin{tikzpicture}
  \draw (0*\constantx,0*\constanty) node(s0) [circle, draw, scale=\nodescale] {}
        (1*\constantx,-1*\constanty) node(s1) [circle, draw, scale=\nodescale] {}
        (1*\constantx,-3*\constanty) node(s2) [circle, draw,scale=\nodescale] {}
        (4*\constantx,-1*\constanty) node(s3) [circle, draw,scale=\nodescale] {}
        (4*\constantx,-3*\constanty) node(s4) [circle, draw,scale=\nodescale] {}
        (8*\constantx,-4*\constanty) node(s5) [circle, draw,scale=\nodescale] {}
        (5*\constantx,-7*\constanty) node(s6) [circle, draw,scale=\nodescale] {}
        (6*\constantx,-5*\constanty) node(s8) [circle, draw,scale=\nodescale] {};
  \draw (1*\constantx,0*\constanty) node(a1) [scale=\letterscale] {$b$}
        (4*\constantx,0*\constanty) node(d4) [scale=\letterscale] {$a$}
        (5*\constantx,0*\constanty) node(c5) [scale=\letterscale] {$c$}
        (6*\constantx,0*\constanty) node(a6) [scale=\letterscale] {$d$}
        (8*\constantx,0*\constanty) node(c5) [scale=\letterscale] {$a$};
  \draw (1*\constantx,0.5*\constanty) node(c1) [scale=\letterfigsc] {$1$}
        (4*\constantx,0.5*\constanty) node(c2) [scale=\letterfigsc] {$i_1$}
        (5*\constantx,0.5*\constanty) node(i3) [scale=\letterfigsc] {$i_1+1$}
        (6*\constantx,0.5*\constanty) node(i4) [scale=\letterfigsc] {$i_1+2$}
        (8*\constantx,0.5*\constanty) node(i5) [scale=\letterfigsc] {$i_2$};
  \draw (-0.6*\constantx,-1*\constanty) node(j1) [scale=\letterfigsc] {$1$}
        (-0.6*\constantx,-3*\constanty) node(j2) [scale=\letterfigsc] {$j_1$}
        (-0.6*\constantx,-4*\constanty) node(j3) [scale=\letterfigsc] {$j_1+1$}
        (-0.6*\constantx,-5*\constanty) node(j4) [scale=\letterfigsc] {$j_1+2$}
        (-0.6*\constantx,-7*\constanty) node(j5) [scale=\letterfigsc] {$j_2$};

 \draw (0*\constantx,-1*\constanty) node(a1) [scale=\letterscale] {$a$}
        (0*\constantx,-3*\constanty) node(d4) [scale=\letterscale] {$b$}
        (0*\constantx,-4*\constanty) node(c5) [scale=\letterscale] {$a$}
        (0*\constantx,-5*\constanty) node(a6) [scale=\letterscale] {$d$}
        (0*\constantx,-7*\constanty) node(c5) [scale=\letterscale] {$c$};

\draw  {(1.5*\constantx,0*\constanty) -- (3.5*\constantx,0*\constanty)};
\draw  {(6.5*\constantx,0*\constanty) -- (7.5*\constantx,0*\constanty)};
\draw  {(0*\constantx,-1.5*\constanty) -- (0*\constantx,-2.5*\constanty)};
\draw  {(0*\constantx,-5.5*\constanty) -- (0*\constantx,-6.5*\constanty)};

  \draw[->] (s0) edge[bend right=15] node[above, scale=\letterfigsc] {$a$} (s3);
  \draw[->] (s0) edge[bend left=15] node[right, scale=\letterfigsc] {$b$} (s2);
  \draw[->] (s0) edge[style=dashed] node[below, scale=\letterfigsc] {} (s1);

  \draw[->] (s4) edge node[above, scale=\letterfigsc] {$a$} (s5);
  \draw[->] (s4) edge node[right, scale=\letterfigsc] {$c$} (s6);
  \draw[->] (s4) edge[bend right=15,style=dashed] node[below, scale=\letterfigsc] {} (s8);
  \draw[->] (s4) edge[bend left=15] node[above, scale=\letterfigsc] {$d$} (s8);

    

  \end{tikzpicture}
\caption{\label{fig:simple2d} An incomplete common subsequence automaton for two strings $S_1,S_2$, laid out in a two-dimensional grid. State $(0,0)$ has a transition labeled $a=S_2[1]$ to state $(i_1, 1)$ and a transition labeled $b=S_1[1]$ to state $(0, j_1)$. State $(i_1, j_1)$ is at level 1 and has a transition for each unique character in $\Sigma(S_1[i_1+1, i_1+2]) \cup \Sigma(S_2[j_1+1, j_1+2])=\{a,c,d\}$. Transitions out of the remaining states are missing from the illustration.}
\end{figure}

An example of an incomplete common subsequence automaton for two strings is given in Figure~\ref{fig:simple2d}.

\subsubsection{Analysis}

For every pair of positive integers $(i,j)$, we have the following property:
\begin{gather*}
\overline{(i,j)} - (i,j) = 2^{\level((i,j))}
\end{gather*}
This property follows from the same argument that led to equation (\ref{sec:overlinesminuss}). The number of transitions out of every state $s$, is now bounded by $2\cdot 2^{\level(s)}$ since both $S_1$ and $S_2$ can contribute with up to $2^{\level(s)}$ transitions.

We can calculate the size of the alphabet-aware level automaton for two strings by summing up the space contribution from each diagonal of states. Let $d$ be a diagonal consisting of $|d|$ states. Then the size of $d$ is $O(|d| \log \sigma)$, since each diagonal has the size of an alphabet-aware level automaton for one string. If $D$ is the set of all diagonals, then the total size of the automaton becomes
\begin{gather*}
\sum_{d\in D} O(|d|\log \sigma)=\log \sigma \cdot \sum_{d\in D} O(|d|)= O(\log \sigma \cdot  n_1 \cdot  n_2 )
\end{gather*}
The last step is possible since the sum over the states in all diagonals is exactly the number of states in the automaton. In summary we have shown the following result:

\begin{lemma}
Let $S_1,S_2$ be strings of length $n_1$ and $n_2$ over an alphabet of size $\sigma$. We can construct a subsequence automaton and a common subsequence automaton with default transitions of size $O(n_1 n_2 \log \sigma)$ and delay $O(\log \sigma)$.
\end{lemma}

\subsection{The alphabet-aware level automaton for multiple strings}
The alphabet-aware level automaton for the set of strings $\mathcal{S}=\{S_1, S_2, \ldots, S_N\}$ have $1+\prod_{i=1}^N S_i$ states that we identify with the set of integer points $\{1,\ldots,  n_1 \}\times \{1,\ldots,  n_2 \}\times\ldots\times\{1,\ldots,  n_N \}\cup \{(0,0,\ldots, 0)\}$. Hence, a state in the automaton corresponds to a tuple with $N$ elements $(s_1,s_2,\ldots, s_N)$. We generalize the definition of diagonals for dimension $N$ as follows. The diagonal of a state $(s_1,s_2, \ldots, s_N)$ is the set of states:
\begin{gather*}
\{(s_1+k, s_2+k,\ldots, s_N+k) \;|\; \bigwedge\limits_{i=1}^N 0 < s_i+k \leq  n_i \}
\end{gather*}
Again, states belong to the same diagonal if the diagonal of each state defines identical sets of states, and for states $(s_1, s_2,\ldots, s_N),(s_1',s_2',\ldots, s_N')$, in the same diagonal, $(s_1, s_2,\ldots, s_N)<(s_1',s_2',\ldots, s_N')$ if $s_1 < s_1'$. For each state $(s_1, s_2, \ldots, s_N)$ we associate the integer $\min(s_1,s_2,\ldots, s_N)$ and define subtraction and modulo operations on states as in the previous section.

With each state we again associate the level:
\begin{gather*}
\level((s_1, s_2, \ldots, s_N)) = \min(\left\lceil \log_2 \sigma \right\rceil, \max(\{x \;|\; (s_1, s_2, \ldots, s_N) \bmod 2^x = 0 \}))
\end{gather*}

For a tuple of positive integers $(s_1,s_2, \ldots, s_N)$, we define $\overline{(s_1,s_2, \ldots, s_N)}>(s_1,s_2, \ldots, s_N)$ to be the smallest tuple of integers in the same diagonal such that $\level((s_1,s_2, \ldots, s_N)) < \level(\overline{(s_1,s_2, \ldots, s_N)})$.

The alphabet-aware level automaton for multiple strings, $S_1,S_2,\ldots, S_N$, has the following transitions: State $(0,0,\ldots,0)$ has a transition labeled $S_i[1]$ to state $(s_1, s_2, \ldots,s_N)$ where $s_i=1$, such that $s_j$ is the minimal index where $S_j[s_j]=S_i[1]$, for every $i=\{1,2, \ldots, N\}$ and $j\not = i$ and a default transition to state $(1,1, \ldots, 1)$.
Every other state $(s_1,s_2, \ldots, s_N)$, where $(\overline{s_1}, \overline{s_2}, \ldots, \overline{s_N})= \overline{(s_1,s_2, \ldots, s_N)}$, has the following transitions:
\begin{itemize}
\item A default transition to state $\overline{(s_1,s_2, \ldots, s_N)}$. If no such state exist, the state $(s_1,s_2, \ldots, s_N)$ does not have a default transition.
\item If $\overline{(s_1,s_2, \ldots, s_N)} - (s_1,s_2, \ldots, s_N) < \sigma$ then for each character $\alpha$ in $\bigcup_{i=1}^N \Sigma(S_i[s_i + 1, \min(\overline{s_i},  n_i )])$  there is a transition labeled $\alpha$ to the state $(s_1',s_2', \ldots, s_N')$, where, $s_i'>s_i$ is the minimal index such that $S_i[s_i']= \alpha$, for all $1\leq i \leq N$
\item If $\overline{(s_1,s_2, \ldots, s_N)} - (s_1,s_2, \ldots, s_N) \geq \sigma$ then for each character $\alpha$ in $\bigcup_{i=1}^N \Sigma(S_i[s_i + 1,  n_i ])$ there is a transition labeled $\alpha$ to the state $(s_1',s_2', \ldots, s_N')$, where $s_i'>s_i$ is the minimal index such that $S_i[s_i'] = \alpha$, for all $1\leq i \leq N$.
\end{itemize}

\subsubsection{Analysis}
For each state $(s_1, s_2, \ldots, s_N)$ we have that 
\begin{gather*}
\overline{(s_1, s_2, \ldots, s_N)} - (s_1, s_2, \ldots, s_N) = 2^{\level((s_1, s_2, \ldots, s_N))}
\end{gather*}

The number of transitions out of every state $s$, is now bounded by $N\cdot 2^{\level(s)}$ because each of the $N$ strings can contribute with up to $2^{\level(s)}$ transitions.

 We can calculate the size of the alphabet-aware level automaton for $N$ strings by summing up the space contribution from each diagonal of states. Let $d$ be a diagonal consisting of $|d|$ states. Then the size of $d$ is $O(N|d| \log \sigma)$. If $D$ is the set of all diagonals, then the total size of the automaton becomes
\begin{gather*}
O\left (\sum_{d\in D} N|d|\log \sigma\right)=O\left (N \log \sigma \cdot \sum_{d\in D} |d|\right)= O \left (N \log \sigma \cdot \prod_{i=1}^N  n_i \right )
\end{gather*}
The last step is possible since the sum over the states in all diagonals is the number of states in the automaton. In summary we have shown the following result:
\begin{theorem}
Let $S_1,S_2, \ldots S_N$ be a set of strings of length $n_1,n_2, \ldots, n_N$ over an alphabet of size $\sigma$. We can construct a subsequence automaton and a common subsequence automaton with default transitions of size $O(N \log \sigma \cdot \prod_{i=1}^N n_i)$ and delay $O(\log \sigma)$.
\end{theorem}





\bibliographystyle{splncs}
\bibliography{references}

\end{document}